\documentclass[usegraphicx]{mn2e}

\usepackage{amssymb}
\usepackage{mathptmx}
\newcommand{\Mpc}{\rm\thinspace Mpc}

\newcommand{\km}{\rm\thinspace km}

\newcommand{\cm}{\rm\thinspace cm}

\newcommand{\pcmcu}{\hbox{$\cm^{-3}\,$}}

\newcommand{\s}{\rm\thinspace s}

\newcommand{\Msun}{\hbox{$\rm\thinspace M_{\odot}$}}

\newcommand{\erg}{\rm\thinspace erg}

\newcommand{\ergpcmsqps}{\hbox{$\erg\cm^{-2}\s^{-1}\,$}}

\newcommand{\ergps}{\hbox{$\erg\s^{-1}\,$}}

\newcommand{\kmps}{\hbox{$\km\s^{-1}\,$}}

\newcommand{\kmpspMpc}{\hbox{$\kmps\Mpc^{-1}$}}

\newcommand{\psqcm}{\hbox{$\cm^{-2}\,$}}
\newcommand{\pcmsq}{\hbox{$\cm^{-2}\,$}}

\begin{document}

\title[AGN in the Perseus cluster]{X-ray active galactic nuclei in the
  core of the Perseus cluster}

\author[S. Santra, J.~S. Sanders \& A.~C. Fabian]
{S. Santra\thanks{E-mail:sumita@ast.cam.ac.uk}, J.~S. Sanders\thanks{E-mail:jss@ast.cam.ac.uk} and
  A.~C. Fabian\\ Institute of Astronomy, Madingley Road, Cambridge. CB3
  0HA}
\maketitle
\begin{abstract}
  We present a study of the X-ray emission from the nuclei of galaxies
  observed in the core of the Perseus cluster in a deep exposure with
  \emph{Chandra}. Point sources are found coincident with the nuclei
  of 13 early-type galaxies, as well as the central galaxy NGC\,1275.
  This corresponds to all galaxies brighter than $M_{\rm B}=-18$ in
  the \emph{Chandra} field.  All of these sources have a steep
  power-law spectral component and four have an additional thermal
  component. The unabsorbed power-law luminosities in the 0.5--7.0~keV
  band range from $8\times10^{38}$--$5\times10^{40} \ergps$. We find
  no simple correlations between the K band luminosity, or the FUV and
  NUV AB magnitudes of these galaxies and their X-ray properties. We
  have estimated the black hole masses of the nuclei using the K band
  M$_{\rm BH}$-L$_{K \rm bol}$ relation and again find no correlation
  between black hole mass and the X-ray luminosity. Bondi accretion
  onto the black holes in the galaxies with mini-haloes should make
  them much more luminous than observed.
  
\end{abstract}


\section{Introduction}

X-ray observations with the \emph{Chandra} observatory have enabled
detailed studies to be made of the X-ray emission of early-type
galaxies. The results show hot diffuse gas and low mass X-ray binaries
(LMXB), many of which are associated with globular clusters (Sarazin
et al 2001; Kim et al 2004a; Fabbiano \& White 2003).  The hot gas has
a temperature of $10^7$ K and is not expected to survive in the cores
of rich clusters of galaxies, where it should be rapidly depleted by
stripping and/or conduction (Gunn \& Gott 1972).  However, small
confined regions, of radius a few kpc, do survive in some early-type
galaxies near the centres of rich clusters, as discovered in
\emph{Chandra} observations of the Coma cluster by Vikhlinin et al
(2001). Yamasaki et al (2002) later found two small galaxy coronae in
the centre of the A1060 cluster. Sun et al (2005a) and Sun et al
(2005b) found four such X-ray minicoronae in A1367 and one in the
Perseus cluster, respectively, and Fujita, Sarazin \& Sivakoff (2006)
found one in A2670. A systematic search for X-ray mini-haloes in 157
early-type galaxies and 22 late-type galaxies in 25 hot, rich nearby
clusters by Sun et al (2007) yielded many more examples. From all
these observations the authors have concluded that the diffuse hot
coronae are stripped from the galaxies in the cores of rich clusters
of galaxies, but mini-haloes remain at the centres of some
galaxies. Such mini haloes have typical sizes of 1-3~kpc, gas
densities of $0.1 \pcmcu$ or more and temperatures of about $10^7$~K.

Other studies include Finoguenov et al (2004) who measured the X-ray
luminosity function of galaxies in the Coma cluster using
\emph{XMM-Newton}, and Finoguenov \& Miniati (2004) who studied the
impact of high pressure cluster environment on the X-ray luminosity of
Coma galaxies.  \emph{Chandra} X-ray observations of galaxies in an
off-center region of the Coma cluster were done by Hornschemeier et al
(2006) to explore the X-ray properties and luminosity function of
normal galaxies. They detected 13 galaxies. The X-ray activity is
suppressed with respect to that of the field, indicating a lower level
of X-ray emission for a given stellar mass.

Some nuclei of nearby early-type galaxies are active but often at a
level well below that expected from Bondi accretion of the hot coronal
gas (Fabian \& Canizares 1988; Di Matteo et al 2000, 2001, 2003;
Pellegrini 2005). There are two possible explanations for the low
luminosities of nearby black holes: (1) any accretion proceeds at
extremely low rates or (2) the accretion occurs at low radiative
efficiencies as predicted, for example, by advection-dominated
accretion flow (ADAF) models (e.g., Rees et al 1982; Narayan \& Yi
1995; Abramowicz et al 1995) and by jet models (e.g. Allen et al 2006).

Here we study the X-ray point sources coincident with member galaxies
in the Perseus cluster using a very deep, 900~ks, \emph{Chandra}
image. The Perseus cluster is the brightest cluster in the Sky in
X-rays, which makes looking for faint or diffuse emission difficult,
particularly in the core where our observations are best. In this
paper we have looked at X-ray point sources coincident with bright
galaxies. We are interested here in the unresolved point sources with
power-law spectra, coincident with the galactic nuclei, in order to
assess the level of nuclear activity. Sun et al (2007) have previously
used the same \emph{Chandra} image of the Perseus cluster, and
\emph{Chandra} images of several other clusters, to explore the
incidence and general properties of mini-haloes in cluster
galaxies. Martini et al (2006) found that 5 per cent of the galaxies
brighter than $M_{\rm B}<-20$ in rich clusters are active, with X-ray
luminosity $L_{\rm X}>10^{41}\ergps$.  This is 5 times more than found
in optical spectroscopic studies by Dressler et al (1985). Our results
are consistent with all galaxies above that optical magnitude limit
having detected central X-ray sources, albeit at lower luminosities.

This paper is organised as follows: In Section 2 we describe the data
preparation, source detection and the X-ray spectral analysis. We
compare the X-ray properties with the K band luminosity, UV AB
magnitude, radio and black hole mass in Section 3.  In Section 4 we
discuss our results and summarise them in Section 5.

\section{Data analysis}
\subsection{Observations}


The 13 \emph{Chandra} observations analysed here are detailed in
Fabian et al (2006). Each observation used the ACIS-S3 detector as the
aimpoint. The total maximum effective exposure time was 890 ks, after
flares were removed. The level 2 event files were reprocessed with
\textsc{acis\_process\_events} using the
acisD2000-01-29gain\_ctiN0003.fits gain file, and reprojected to match
the coordinate system of the 04952 observation.

\begin{figure*}
  \includegraphics[width=\textwidth]{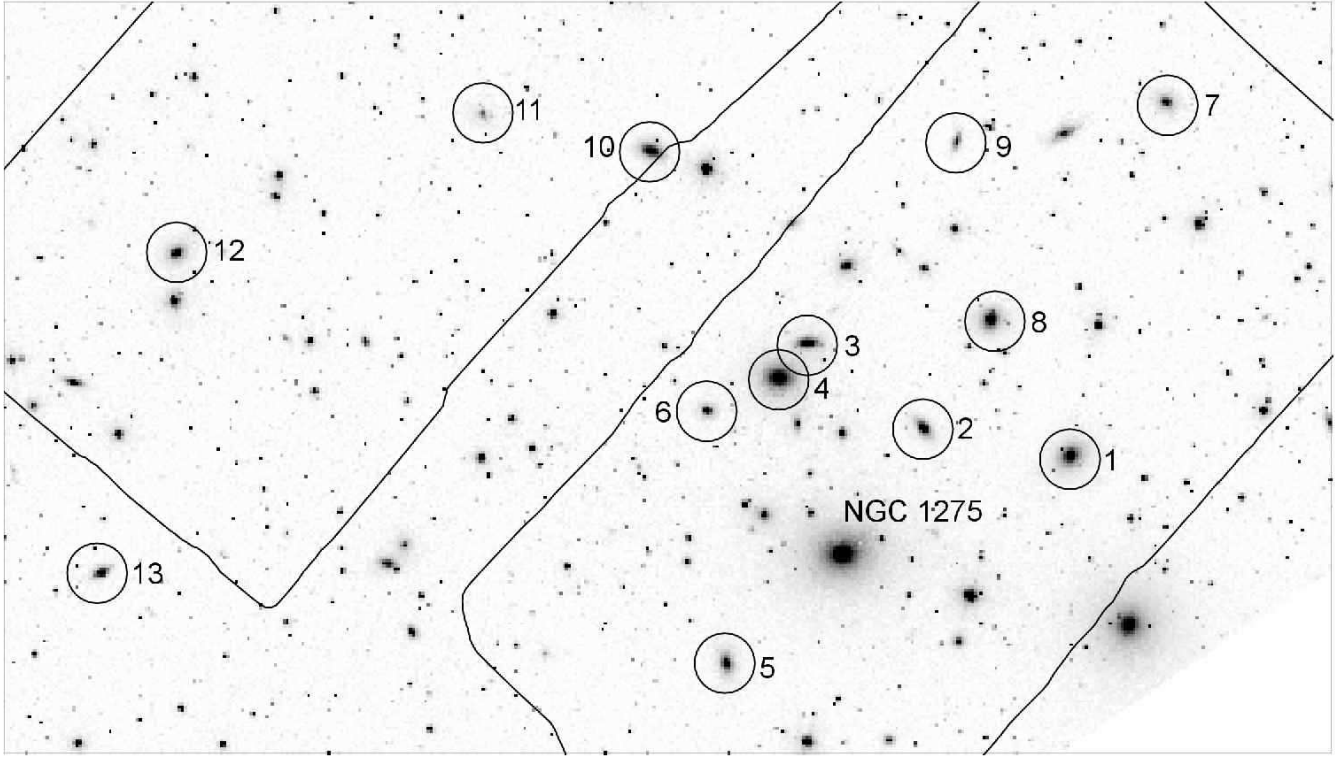}
  \caption{Optical images of the detected X-ray sources from SDSS
    (numbered according to Table~\ref{tab:position}). The two regions
    with greater than 25 per cent of the maximum \emph{Chandra}
    exposure time are shown by the black contour (each region measures
    8~arcmin across).}
  \label{fig:picture4}
\end{figure*}

The regions of the sky with the highest effective \emph{Chandra}
exposure time are shown in Fig.~\ref{fig:picture4}. The off-axis
effective exposure decreases away from the centre of the ACIS-S3 CCD
(which is the SW part of the right region, with a total exposure time
of 890 ks).  In the NE and NW part of the right chip the exposure
times are 75 per cent and 50 per cent of the maximum respectively. The
exposure time declines to 13 per~cent of the maximum in the region
indicated by source 13.

\subsection{X-ray source detection}

\begin{table}
  \caption{Point sources obtained from the \emph{Chandra} data which
    haven an optical counterpart. The sources are split into two
    groups, those on the central CCD, and those on other CCDs. Each
    section is sorted in RA. Positions use J2000 coordinates.}
\begin{tabular}{llll} \hline
N & RA&Dec & Nearest galaxy name\\
&&&from NED (J200)\\
\hline
1 & $03^\mathrm{h}19^\mathrm{m}26.73^\mathrm{s}$ & $+41^\circ32' \: 26.0''$& NGC 1273\\
2 & $03^\mathrm{h}19^\mathrm{m}40.57^\mathrm{s}$ & $+41^\circ32' \: 55.2''$ &NGC 1274\\
3& $03^\mathrm{h}19^\mathrm{m}51.50^\mathrm{s}$ & $+41^\circ34' \: 24.8''$ &NGC 1277\\
4 &$03^\mathrm{h}19^\mathrm{m}54.13^\mathrm{s}$ & $+41^\circ33' \: 48.4''$ &NGC 1278\\
5&$03^\mathrm{h}19^\mathrm{m}59.05^\mathrm{s}$ & $+41^\circ28' \: 46.5''$ &NGC 1279\\
6&$03^\mathrm{h}20^\mathrm{m}00.91^\mathrm{s}$ & $+41^\circ33' \: 13.8''$&Vzw 339\\ \hline
7 &$ 03^\mathrm{h}19^\mathrm{m}17.68^\mathrm{s}$ & $+41^\circ38' \: 37.8''$& 2MASX J03191772+4138391\\
8&$03^\mathrm{h}19^\mathrm{m}34.22^\mathrm{s}$ & $+41^\circ34' \: 50.0''$&CGCG 540-101\\
9&$03^\mathrm{h}19^\mathrm{m}37.38^\mathrm{s}$ & $+41^\circ37' \: 58.9''$ &2MASX J03193743+4137580\\
10&$03^\mathrm{h}20^\mathrm{m}06.19^\mathrm{s}$ & $+41^\circ37' \: 46.3''$ &NGC 1281 \\
11&$ 03^\mathrm{h}20^\mathrm{m}21.42^\mathrm{s}$ & $+41^\circ38' \: 23.9''$ &MCG+07-07-070\\
12&$03^\mathrm{h}20^\mathrm{m}50.60^\mathrm{s}$ & $+41^\circ35' \: 57.7''$ &2MASX J03205074+4136015\\
13&$03^\mathrm{h}20^\mathrm{m}57.71^\mathrm{s}$ & $+41^\circ30' \: 20.4''$ &2MASX J03205776+4130229\\
\hline
\end{tabular}
\label{tab:position}
\end{table}

\begin{table*}
  \caption{Point sources excluded from further analysis since they have no 
    cluster galaxy counterpart. Fluxes are in $\ergpcmsqps$ in the
    0.5-7 keV band. They were measured
    by fitting absorbed powerlaws and \textsc{mekal} components (when
    necessary) to the spectra in the 0.5 to 7~keV band. }
 \begin{minipage}[t]{0.33\textwidth}
 \begin{tabular}{lll}
    \hline
    RA & Dec & Flux\\ \hline
$03^\mathrm{h}19^\mathrm{m}09.52^\mathrm{s}$ & $+41^\circ 34' \: 29.2''$ & $1.1\times 10^{-14}$\\
$03^\mathrm{h}19^\mathrm{m}11.97^\mathrm{s}$ & $+41^\circ 33' \: 53.1''$ & $4.5\times 10^{-15}$\\
$03^\mathrm{h}19^\mathrm{m}15.76^\mathrm{s}$ & $+41^\circ 34' \: 38.3''$ & $3.2\times 10^{-15}$\\
$03^\mathrm{h}19^\mathrm{m}15.95^\mathrm{s}$ & $+41^\circ 34' \: 07.9''$ & $7.8\times 10^{-16}$\\
$03^\mathrm{h}19^\mathrm{m}19.93^\mathrm{s}$ & $+41^\circ 31' \: 56.2''$ & $5.0\times 10^{-15}$\\
$03^\mathrm{h}19^\mathrm{m}20.10^\mathrm{s}$ & $+41^\circ 37' \: 46.4''$ & $2.1\times 10^{-14}$\\
$03^\mathrm{h}19^\mathrm{m}21.12^\mathrm{s}$ & $+41^\circ 31' \: 19.4''$ & $6.2\times 10^{-14}$\\
$03^\mathrm{h}19^\mathrm{m}22.26^\mathrm{s}$ & $+41^\circ 39' \: 37.3''$ & $4.4\times 10^{-15}$\\
$03^\mathrm{h}19^\mathrm{m}23.64^\mathrm{s}$ & $+41^\circ 37' \: 08.7''$ & $3.6\times 10^{-14}$\\
$03^\mathrm{h}19^\mathrm{m}25.00^\mathrm{s}$ & $+41^\circ 31' \: 32.2''$ & $1.2\times 10^{-15}$\\
$03^\mathrm{h}19^\mathrm{m}25.13^\mathrm{s}$ & $+41^\circ 40' \: 26.5''$ & $1.3\times 10^{-14}$\\
$03^\mathrm{h}19^\mathrm{m}28.11^\mathrm{s}$ & $+41^\circ 34' \: 25.4''$ & $3.0\times 10^{-15}$\\
\hline
       \end{tabular}
 \end{minipage}
\begin{minipage}[t]{0.33\textwidth}
  \begin{tabular}{lll}
    \hline
    RA & Dec &Flux\\ \hline
$03^\mathrm{h}19^\mathrm{m}30.02^\mathrm{s}$ & $+41^\circ 40' \: 12.3''$ &$2.3\times 10^{-14}$\\
$03^\mathrm{h}19^\mathrm{m}30.69^\mathrm{s}$ & $+41^\circ 35' \: 04.1''$ & $9.5\times 10^{-15}$\\
$03^\mathrm{h}19^\mathrm{m}33.81^\mathrm{s}$ & $+41^\circ 29' \: 55.9''$ & $3.3\times 10^{-15}$\\
$03^\mathrm{h}19^\mathrm{m}35.67^\mathrm{s}$ & $+41^\circ 27' \: 49.9''$ & $1.7\times 10^{-15}$\\
$03^\mathrm{h}19^\mathrm{m}37.49^\mathrm{s}$ & $+41^\circ 38' \: 44.2''$ &$2.0\times 10^{-14}$\\
$03^\mathrm{h}19^\mathrm{m}38.37^\mathrm{s}$ & $+41^\circ 27' \: 53.5''$ & $1.3\times 10^{-15}$\\
$03^\mathrm{h}19^\mathrm{m}43.68^\mathrm{s}$ & $+41^\circ 27' \: 25.1''$ & $2.2\times 10^{-15}$\\
$03^\mathrm{h}19^\mathrm{m}43.91^\mathrm{s}$ & $+41^\circ 33' \: 04.7''$ & $5.2\times 10^{-15}$\\
$03^\mathrm{h}19^\mathrm{m}44.10^\mathrm{s}$ & $+41^\circ 25' \: 53.2''$ & $1.6\times 10^{-14}$\\
$03^\mathrm{h}19^\mathrm{m}45.47^\mathrm{s}$ & $+41^\circ 42' \: 19.5''$ &$6.0 \times 10^{-15}$\\
$03^\mathrm{h}19^\mathrm{m}46.27^\mathrm{s}$ & $+41^\circ 37' \: 35.7''$ &$7.4\times 10^{-14}$\\
$03^\mathrm{h}19^\mathrm{m}47.75^\mathrm{s}$ & $+41^\circ 27' \: 23.5''$ & $7.7\times 10^{-16}$\\
\hline
       \end{tabular}
  \end{minipage}
  \begin{minipage}[t]{0.33\textwidth}
  \begin{tabular}{lll}
    \hline
    RA & Dec &Flux\\ \hline
$03^\mathrm{h}19^\mathrm{m}48.24^\mathrm{s}$ & $+41^\circ 32' \: 49.7''$ & $9.6\times 10^{-16}$\\
$03^\mathrm{h}19^\mathrm{m}51.44^\mathrm{s}$ & $+41^\circ 31' \: 51.0''$ & $3.7\times 10^{-15}$\\
$03^\mathrm{h}19^\mathrm{m}56.09^\mathrm{s}$ & $+41^\circ 33' \: 15.4''$ & $1.3\times 10^{-14}$\\
$03^\mathrm{h}19^\mathrm{m}59.91^\mathrm{s}$ & $+41^\circ 39' \: 36.1''$ &$5.1\times 10^{-15}$\\
$03^\mathrm{h}20^\mathrm{m}01.50^\mathrm{s}$ & $+41^\circ 31' \: 27.5''$ &$2.5\times 10^{-14}$\\
$03^\mathrm{h}20^\mathrm{m}02.71^\mathrm{s}$ & $+41^\circ 30' \: 33.3''$ & $9.0\times 10^{-16}$\\
$03^\mathrm{h}20^\mathrm{m}04.76^\mathrm{s}$ & $+41^\circ 38' \: 39.5''$ &$5.1 \times 10^{-15}$\\
$03^\mathrm{h}20^\mathrm{m}05.34^\mathrm{s}$ & $+41^\circ 30' \: 54.5''$ & $4.3\times 10^{-15}$\\
$03^\mathrm{h}20^\mathrm{m}11.52^\mathrm{s}$ & $+41^\circ 36' \: 59.4''$ &$6.1 \times 10^{-15}$\\
$03^\mathrm{h}20^\mathrm{m}14.31^\mathrm{s}$ & $+41^\circ 36' \: 05.8''$ & $4.8\times 10^{-15}$\\
$03^\mathrm{h}20^\mathrm{m}18.24^\mathrm{s}$ & $+41^\circ 37' \: 24.0''$ & $2.8\times 10^{-15}$\\
$03^\mathrm{h}20^\mathrm{m}19.69^\mathrm{s}$ & $+41^\circ 37' \: 30.8''$ & $5.8\times 10^{-15}$\\
 \hline
  \end{tabular}
  \end{minipage}
\label{tab:position2}
\end{table*}

We found 49 X-ray sources by eye in the total \emph{Chandra} 0.3-7~keV
image. Of those sources, 13 have optical counterparts. The positions
of these 13 sources and the nearest galaxy are listed in
Table~\ref{tab:position}. The sources are split into six close to the
central galaxy (detected on the central CCD) and seven further away
(detected by other CCDs). The remaining 36 sources without optical
counterparts are shown in Table~\ref{tab:position2}. We exclude these
sources from the remainder of our analysis.

We detected X-ray sources for all the galaxies brighter than $M_{\rm
  B}<-18$. In this deep observation one might expect to detect
galaxies fainter than $M_{\rm B}>-18$, but the diffuse cluster
emission makes the background too high for such work.

The red Digitized Sky Survey (DSS2) was first used to identify
possible optical counterparts for the X-ray sources. We then examined
archival images from the Sloan Digital Sky Survey (SDSS) and the
catalogue of Perseus cluster galaxies by Brunzendorf \& Meusinger
(1999).  Fig.~\ref{fig:picture4} shows the SDSS I band image, indicating
the detected X-ray sources with optical counterparts. They are all
listed as early-type galaxies in the work of Brunzendorf \& Meusinger
(1999). We do not detect a three-armed spiral galaxy (UGC 2665, listed
in Brunzendorf \& Meusinger 1999), which lies between sources 7 and 9
(Fig.~\ref{fig:picture4}).  This morphology excludes it being an
early-type galaxy. Moreover, the velocity of this galaxy is $2800
\kmps$ higher than for the Perseus cluster core, so it is likely an
outlier.

\begin{figure}
  \includegraphics[width=\columnwidth] {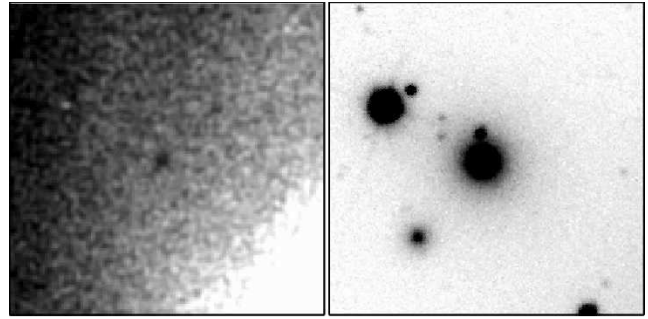}
  \caption{0.3--1.2~keV X-ray image of PGC\,12443 showing drop in
    surface brightness due to absorption (left) compared with SDSS
    I-band image (right). The images measure 1 arcmin vertically.}
  \label{fig:absgal}
\end{figure}

\begin{figure*}
  \includegraphics[width=\textwidth]{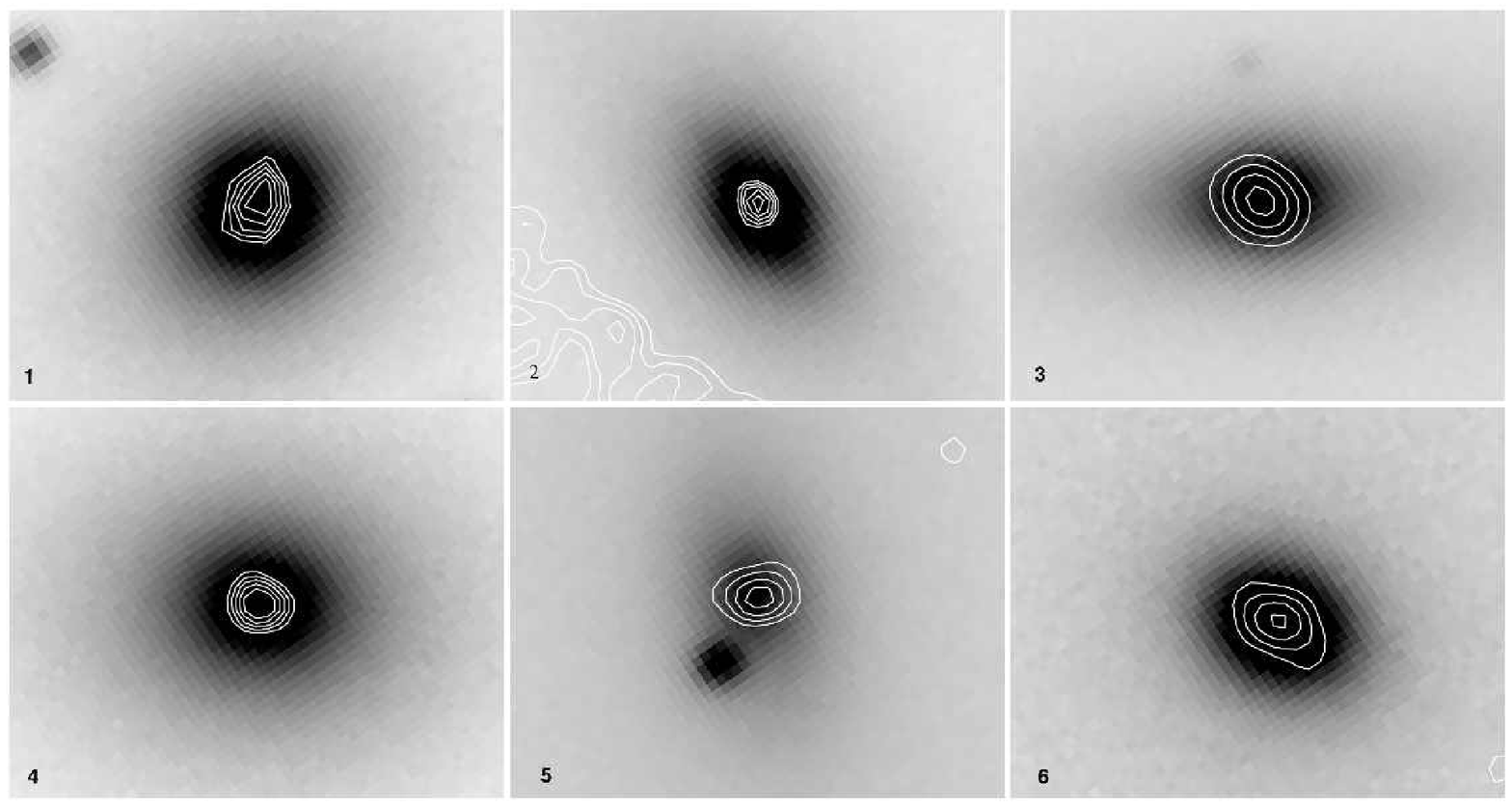}
  \includegraphics[width=\textwidth]{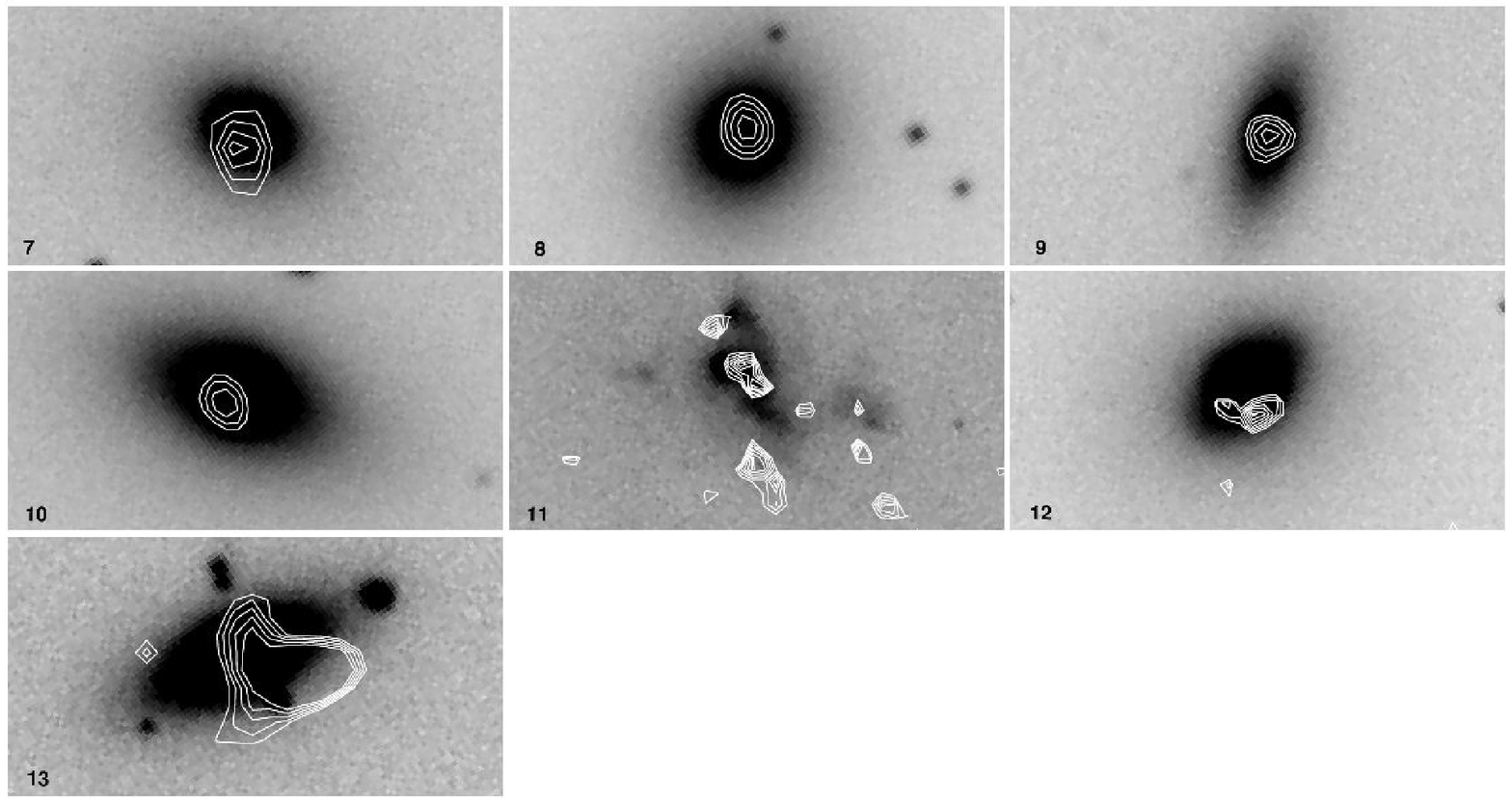}
  \caption{X-ray contours for each of the sources with optical
    counterparts, overlaid on the SDSS I band optical images. The
    first six sources are nearer to the central galaxy and the rest
    are further away.  The first six images are
    27.8$\arcsec$$\times$16.5$\arcsec$ in size, the next seven are
    55.2$\arcsec$$\times$23$\arcsec$.}
  \label{fig:source}
\end{figure*}

In Fig.~\ref{fig:source} we show the X-ray contours for each of the
detected sources (from a 0.3--7 keV image), overlaid on optical images
of the galaxies from SDSS. The first six sources are nearer to NGC
1275 than the others. It is seen that in each of these cases the
contours coincide with the nucleus of the galaxy, so the X-ray sources
match the optical sources. They are also consistent with being point
sources. The remaining seven sources are further from the central
galaxy. Some of these distant sources appear off-centred with respect
to the optical galaxy centres. However, the offsets for these weak
sources are only about 2~arcsec, similar to the offsets obtained for
\emph{Chandra} off-axis sources by Kim et al (2004b). They are mainly
due to the off-axis point spread function (PSF) having an complex
extended shape.  The count rate and distance of the galaxies from the
central galaxy are given in Table~\ref{tab:fittedvalue}.

One galaxy is seen in absorption (Fig.~2). This is listed as 284 or
PGC\,12443 in the Table of Brunzendorf \& Meusinger (1999). It lies
midway between galaxies 4 and 5 in Fig.~1 and is centred at $03^{\rm
  h}19^{\rm m}55.6^{\rm s}$, $+41^\circ31'23''$ and is listed as an
elliptical galaxy 1.9 magnitudes fainter than NGC\,1278. A spectral
fit shows that the absorption corresponds to an intrinsic column
density of $\sim 3\pm 1 \times 10^{20}\psqcm$.
  
\subsection{X-ray spectral analysis}
We extracted and modeled the X-ray spectra of those sources identified
as Perseus core galaxies. For those sources on the central CCD, we
used an extraction region of 3~arcsec. This increased to 5-10~arcsec
on the other chips. Spectra for each source were extracted from each
of the separate observations and added together. The local backgrounds
were extracted from annuli with outer radii of 5 arcsec on the central
CCD, increasing to 10-20 arcsec on the other chips. Response matrix
files (RMF) and ancillary response files (ARF) were created for each
of the sources by averaging the RMFs and ARFs from the individual
observations (created using \textsc{mkacisrmf} and \textsc{mkwarf}).

The spectral fitting was performed using \textsc{xspec} v11.3.2. The
spectra were grouped to have a minimum of 20 counts in each spectral
bin before background subtraction in order to use the $\chi^2$
statistic. Model fitting was carried out in the $0.5-7.0$ keV band.

Each X-ray spectrum was fitted by a power-law with possible intrinsic
absorption and then, separately, by a {\sc mekal} thermal
spectrum. Most were best fitted by the power-law spectrum. We let the
power-law slope ($\Gamma$) and the absorbing column density
($N_{\mathrm H}$) vary freely. In some cases the column density was
less than $1.3\times10^{21} \psqcm$ (the galactic absorption value)
which was unphysical, so we fixed it at this minimum. The photon
indexes of sources 5, 8 and 9 were fixed at 2.0. In the cases where a
single component power-law did not give a satisfactory fit, a
\textsc{mekal} component with a temperature fixed at 0.6 keV was added
to improve the fit. We assume that this is a mini-halo component,
although it is not spatially resolved.

The results of the spectral fitting of all the galaxies are given in
Table~\ref{tab:fittedvalue}. The values of $N_{\mathrm H}$ and
$\Gamma$ vary between $(1.3-3.7) \times 10^{21} \psqcm$ and
$(2.0-3.77)$ respectively. The values of column density was fixed at
the minimum for sources 1, 2, 6, 9, 10, 11, 12 and 13. The $\Gamma$
values are much steeper than for LMXB (Irwin, Athey \& Bregman 2003).

\begin{figure}
  \includegraphics[width=\columnwidth]{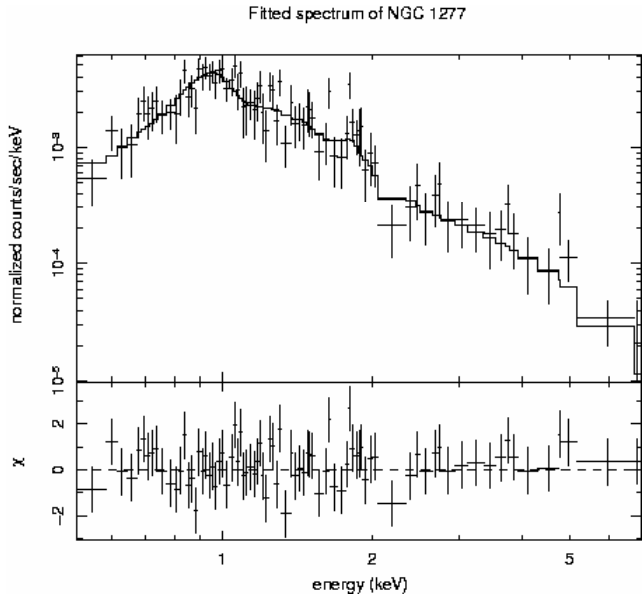}
  \caption{The X-ray spectrum of source 3 showing the best-fit
    absorbed power-law + \textsc{mekal} (solid line). The fit
    residuals are presented in the lower panel.}
  \label{fig:ngc1277}
\end{figure}

A typical spectral fit (for source 3) is shown in
Fig.~\ref{fig:ngc1277}. The spectrum was fitted using an absorbed
power-law and \textsc{mekal} model.

In Table~\ref{tab:Luminosities}, the $0.5-7.0$ keV flux and the
absorption corrected luminosities of the individual sources based on
the best-fitting absorbed power-law model, and in some cases
\textsc{mekal}+absorbed power-law model, are presented. Luminosities
were determined assuming a redshift 0.018 and cosmological model with
$H_0 = 70 \kmpspMpc$. The distances of all of these
galaxies are assumed to be the same as that of NGC\, 1275, as they are
in the same cluster (Brunzendorf \& Meusinger 1999). The uncertainties
on the X-ray luminosities were determined by generating a Monte Carlo
Markov Chain using the built-in \textsc{xspec} functionality. After
the chain had converged, we calculated the luminosity (without
absorption) from each set of values in the chain.  The quoted
luminosity is the median luminosity from the chain, and the
uncertainties were calculated using the 15.85 and 84.15
percentiles. The rest-frame $0.5-2$ keV luminosities of thermal
minihaloes of Perseus core galaxies from Sun et al (2007) are also
given in Table~\ref{tab:Luminosities}. These are the total
(power-law+\textsc{mekal}) luminosities. Sun et al (2007) give no data
for sources 6, 7, 9, 11, 12, 13 and found only two minihalo
objects. We found two more. The absolute total B band magnitudes of
the galaxies are also shown in Table~\ref{tab:Luminosities},
calculated using the luminosity distance and the total B band flux
density from NED. There are no reported B band fluxes for sources 5,
6, 11, 12, 13.

As the first 200~ks of the observations occurred around 2~yr before
the remainder, we investigated whether the detected sources were time
variable. We determined the X-ray flux of the sources for the two
epochs of observation and found the fluxes were the same within the
uncertainties. Therefore no significant variability was found.

\begin{table*}
  \caption {Count rate in the energy range between $0.5-7.0$ keV for
    the detected galaxies, distance from the central galaxy and
    best-fitting parameters of the X-ray spectral fits.}
\label{tab:fittedvalue}
\begin{tabular}{llllllll}
\hline
N&Count rate&Distance from the&Model&$\chi^2$/d.o.f&$N_{\mathrm H}$&$\Gamma$&kT\\
&(counts ks$^{-1}$)&central galaxy (arcmin)&&&($10^{22} \pcmsq$)&&(keV)\\
\hline                                                  
1&$4.99\pm0.08$&4.4&Power-law&$167.64/144$&$0.13$ (fixed)&$2.34^{+0.16}_{-0.16}$&\\
&&&&&&&\\
2&$1.94\pm0.05$&2.6&Power-law&$53.08/66$&$0.13$ (fixed)&$2.37^{+0.31}_{-0.28}$&\\
&&&&&&&\\
3&$9.12\pm0.24$&3.7&Mekal+&$190.57/178$&$0.30^{+0.04}_{-0.04}$&$2.51^{+0.22}_{-0.20}$&0.6\\
&&&Power-law&&&&\\
4&$2.87\pm0.07$&3.3&Power-law&$89.79/95$&$0.19^{+0.06}_{-0.07}$&$2.43^{+0.44}_{-0.25}$&\\
&&&&&&&\\
5&$4.97\pm0.11$&2.8&Mekal+&$170.40/147$&$0.37^{+0.14}_{-0.14}$&$2.0$ (fixed)&$0.6$\\
&&&Power-law&&&&\\
6&$2.48\pm0.06$&3.5& Power-law&$95.46/84$&$0.13$ (fixed)&$2.19^{+0.24}_{-0.22}$&\\
&&&&&&&\\
7&$4.10\pm0.07$&9.8&Power-law&$124.42/124$&$0.24^{+0.12}_{-0.13}$&$3.29^{+0.89}_{-0.56}$&\\
&&&&&&&\\
8&$2.97\pm0.06$&4.9&Mekal+&$94.28/96$&$0.24^{+0.07}_{-0.07}$&$2.0$ (fixed)&$0.6$\\
&&&Power-law&&&&\\
9&$1.63\pm0.04$&7.5&Power-law&$61.4/56$&$0.13$ (fixed)&$2.0$ (fixed)&\\
&&&&&&&\\
10&$1.35\pm0.04$&7.8&Power-law&$30.21/46$&$0.13$ (fixed)&$2.02^{+0.22}_{-0.20}$&\\
&&&&&&&\\
11&$14.89\pm0.13$&9.9&Mekal+&$252.16/268$&$0.13$ (fixed)&$3.25^{+0.66}_{-1.26}$&0.6\\
&&&Power-law&&&&\\
12&$3.21\pm0.06$&12.8&Power-law&$106.28/105$&$0.13$ (fixed)&$3.77^{+0.41}_{-0.38}$&\\
&&&&&&&\\
13&$3.93\pm0.07$&12.98&Power-law&$106.15/124$&$0.13$ (fixed)&$2.43^{+0.17}_{-0.16}$&\\
&&&&&&&\\
\hline
\end{tabular}
\end{table*}

\begin{table*}
  \caption {Total fluxes, velocities, absorption corrected luminosities, 
    the rest frame $0.5-2$ keV luminosities from Sun et al (2007) and total B 
    band magnitude of the sources.}
\label{tab:Luminosities}
\begin{tabular}{lllllll}
\hline
&&&\multicolumn{2}{c|}{}&&\\
N&Total Flux&Velocity&\multicolumn{2}{c|}{Absorption corrected}&Rest frame ($0.5-2$ keV)&Total B band\\ 
&($0.5 -7.0$ keV)&(from NED)&\multicolumn{2}{c|}{ Luminosity ($0.5-7.0$keV)}&luminosities &luminosity\\
&(\ergpcmsqps)&(\kmps)&\multicolumn{2}{c|}{(\ergps)}&(\ergps)&(mag)\\
\cline{4-5}
&&&Mekal&Power-law&(from Sun et al. 2007)&\\
\hline
&&&&&&\\
1&$7.66\times 10^{-15}$&5387&&$7.29^{+0.58}_{-0.57}\times 10^{39}$&$<3.80\times 10^{39}$&-20.04\\
&&&&&&\\
2&$1.90\times 10^{-15}$&6413&&$1.83^{+0.30}_{-0.27}\times 10^{39}$&$<3.16\times 10^{39}$&-19.58\\
&&&&&&\\
3&$1.57\times 10^{-14}$&5066&$6.90^{+5.85}_{-2.72}\times 10^{39}$&$1.69^{+0.23}_{-0.18}\times 10^{40}$&$4.47\times 10^{39}$&-19.51\\
&&&&&&\\
4&$4.26\times 10^{-15}$&6090&&$4.53^{+0.79}_{-1.42}\times 10^{39}$&$<2.75\times 10^{39}$&-21.01\\
&&&&&&\\
5&$2.19\times 10^{-15}$&7285&$2.15^{+0.50}_{-0.47}\times 10^{39}$&$1.22^{+0.72}_{-0.64}\times 10^{39}$&$<3.09\times 10^{39}$&\\
&&&&&&\\
6&$3.36\times 10^{-15}$&5186&&$3.10^{+0.40}_{-0.38}\times 10^{39}$&&\\
&&&&&&\\
7&$3.95\times 10^{-15}$&6211&&$6.17^{+0.56}_{-0.54}\times 10^{39}$&&-18.68\\
&&&&&&\\
8&$3.49\times 10^{-15}$&4500&$4.08^{+1.97}_{-1.33}\times 10^{39}$&$1.09^{+0.70}_{-0.59}\times 10^{39}$&$3.16\times 10^{39}$&-18.48\\
&&&&&&\\
9&$9.09\times 10^{-16}$&8574&&$8.07^{+3.42}_{-3.26}\times 10^{38}$&&-19.56\\
&&&&&&\\
10&$6.55\times 10^{-15}$&4300&&$5.83^{+0.57}_{-0.53}\times 10^{39}$&$<4.37\times 10^{39}$&-19.29\\
&&&&&&\\
11&$7.8\times 10^{-15}$&3749&$5.50^{+1.36}_{-1.60}\times 10^{39}$&$3.66^{+2.70}_{-2.18}\times 10^{39}$&&\\
&&&&&&\\
12&$4.45\times 10^{-15}$&5306&&$5.6^{+0.79}_{-0.77}\times 10^{39}$&&\\
&&&&&&\\
13&$4.82\times 10^{-14}$&4965&&$4.68^{+0.31}_{-0.31}\times 10^{40}$&&\\
&&&&&&\\
\hline
\end{tabular}
\end{table*}

\section{Properties of the early-type galaxies}
We now make a systematic investigation of the properties of the
detected 13 early-type galaxies.

\subsection{K band luminosity and X-ray luminosity}
\begin{figure}
  \includegraphics[width=\columnwidth] {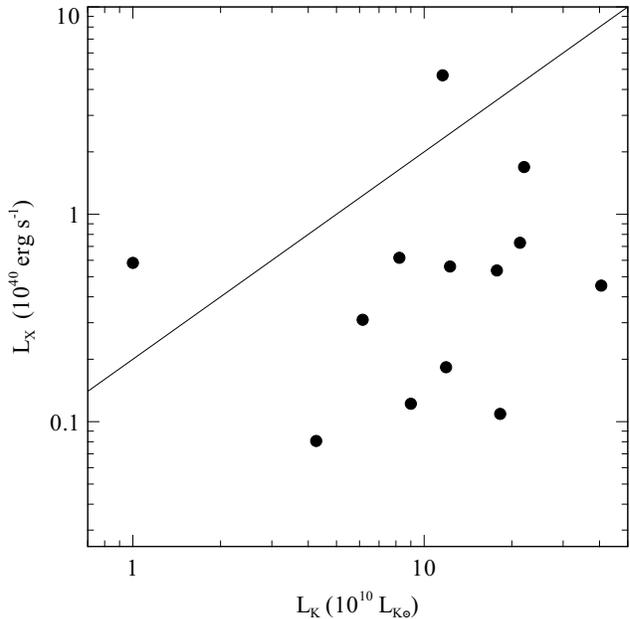}
  \caption{X-ray luminosity vs K band luminosity plot. The straight
    line is the linear relation between the X-ray luminosity of LMXB
    and the K band luminosity of a galaxy, taken from Kim \& Fabbiano
    (2004).}
  \label{fig:Kbandlumin}
\end{figure}

In Fig.~\ref{fig:Kbandlumin} we plot for each source the unabsorbed
power-law X-ray luminosity versus the K band luminosity. L$_K$ was
derived from K$_{20}$ measured within the 20 mag arcsec$^{-2}$
isophote (taken from Two Micron All Sky Survey, 2MASS), using
M$_{K\odot}$ = 3.33 mag.  We also plot a straight line showing the
linear relation between the X-ray luminosity of LMXB in a galaxy and
the galaxy K band luminosity (from Kim \& Fabbiano 2004). Kim \&
Fabbiano (2004) find a range in K band luminosity of 7 to $40\times
10^{10} L_{K\odot}$, which is a similar range to our galaxies.  All
but two of our sources lie below the relation. Therefore the galaxies
appear superficially underluminous in point sources.

\begin{figure*}
  \includegraphics[width=0.8\textwidth] {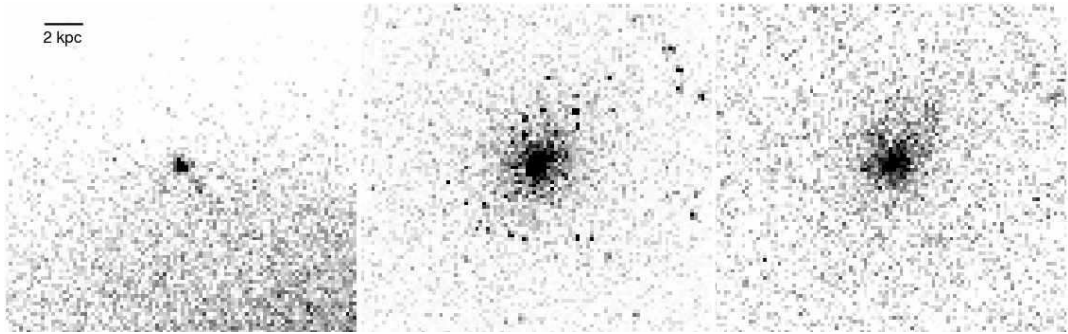}
  \caption{(Left) NGC\,1278 in the Perseus cluster. (Centre) NGC\,720
    at the same physical scale as NGC\,1278. (Right) Simulated image
    of NGC\,720 accounting for projected cluster emission, PSF and
    Poisson noise.}
  \label{fig:fake}
\end{figure*}

To assess the effect of instrumental effects and projection of
intracluster medium on the detectability of point sources, we compared
from Perseus NGC\,1278 ($L_K=40.55 \times 10^{10}L_{K\odot}$), the
optically brightest galaxy detected here, against the
publicly-available \emph{Chandra} observation of NGC\,720 ($L_K=18.58
\times 10^{10}L_{K\odot}$). NGC\,720 lies at a distance of 28 Mpc, is
not in a cluster and has a \emph{Chandra} observation of 40~ks length
(OBSID 492).  To account for the effect of distance on the spatial
scale, we binned the image of NGC\,720 by a factor of 3, and smoothed
it by 1~arcsec to account for the PSF at the position of
NGC\,1278. From this image we subtracted a flat background, then
multiplied the image by a factor of 2.3 to account for the difference
in exposure time and distance between the Perseus and NGC\,720
observations. We added a flat background component to account for the
projected intracluster medium in Perseus. Finally we generated an
image by making a Poisson realization of our model, to account for
counting statistics.  We show an image of NGC\,1278, NGC\,720 before
processing, and the final simulated NGC\,720 image in
Fig.~\ref{fig:fake}.


It is clear that most of the halo of point sources (LMXB) spread
several arcmin around NGC\,720 become undetectable under the observing
conditions of NGC\,1278. The situation would become worse after the
higher absorption to the Perseus cluster is included. We only detect
in NGC\,1278 a point source coincident with the nucleus and a possible
LMXB to the SW. It would be interesting to see whether this last
source coincides with a globular cluster there but we have found no
published information of globular clusters in our 13 galaxies. We
conclude that the low X-ray luminosity of most of the 13 galaxies,
when compared with nearby ones, is due to our inability to resolve the
expected, spatially-distributed, population of point sources.

\subsection{UV band magnitudes, radio emission, black hole mass and X-ray 
luminosity}
\begin{figure*}
  \includegraphics[width=0.45\textwidth]{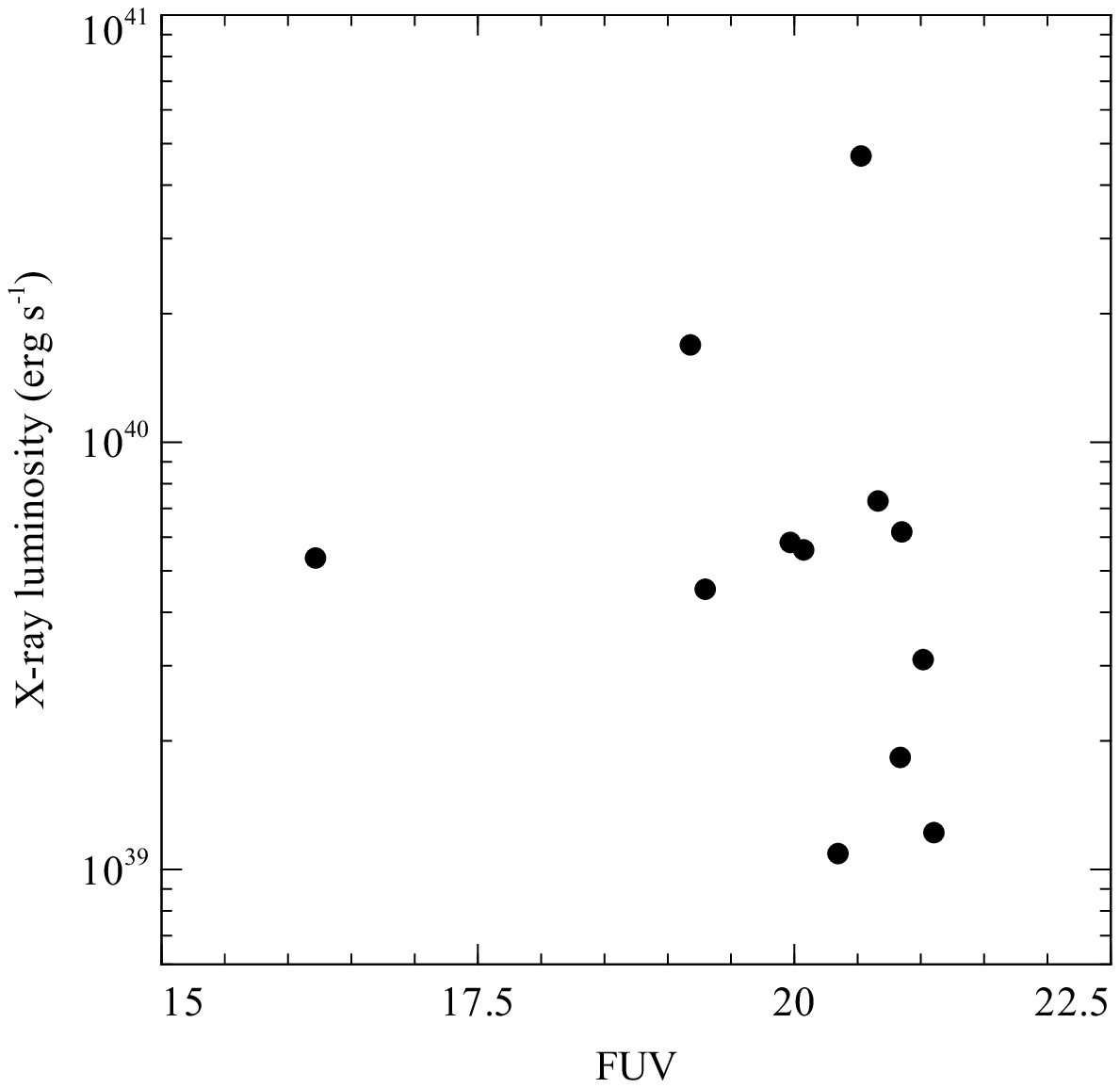}
  \hspace{3mm}
  \includegraphics[width=0.45\textwidth]{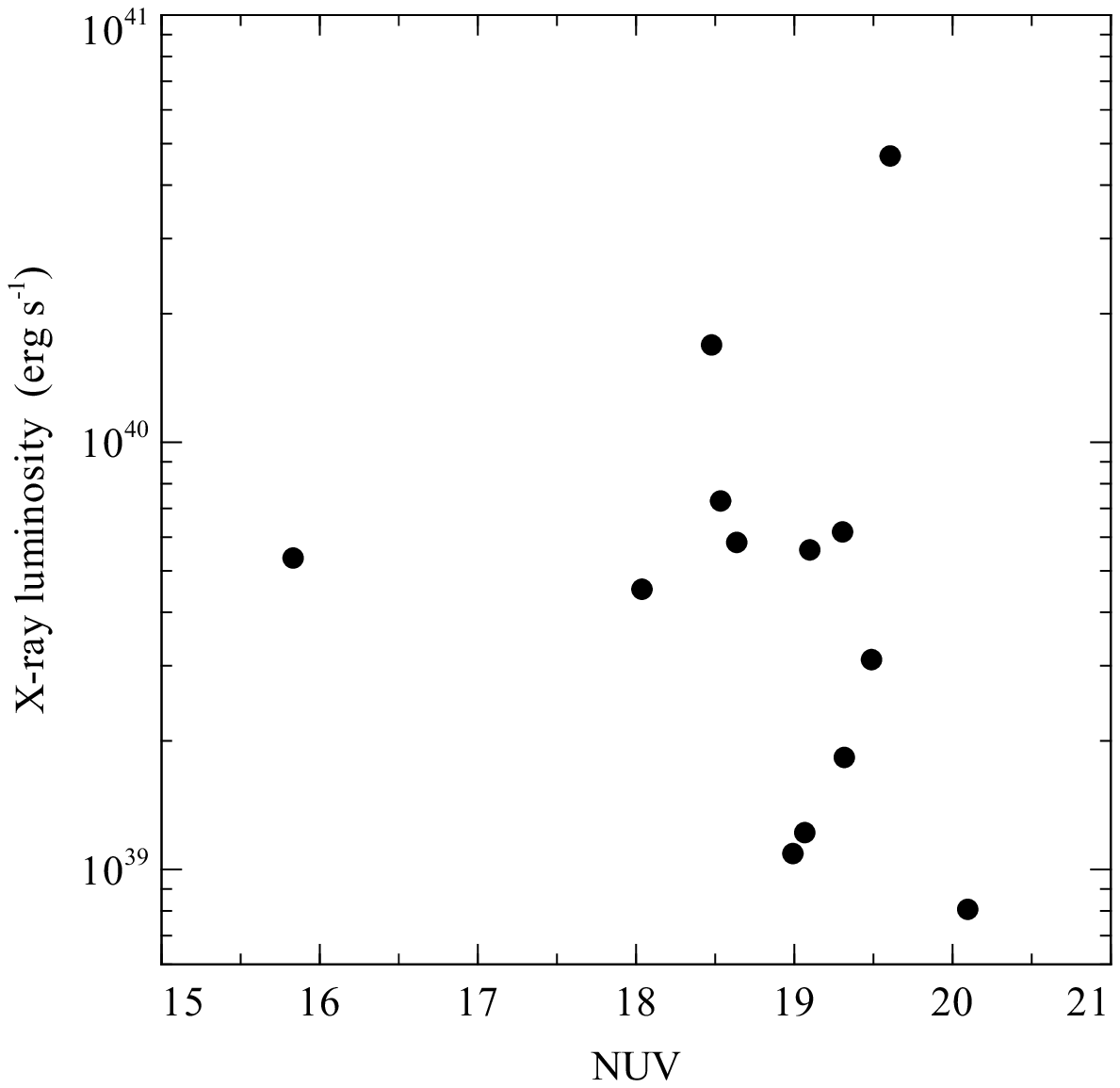}
  \caption{Source FUV (left) and NUV (right) AB magnitudes plotted
    against X-ray luminosity. Note that all sources are essentially at
    the same distance.}
  \label{fig:uvbandmag}
\end{figure*}

\emph{GALEX} FUV and NUV photometry of the centres of the detected
Perseus cluster galaxies was taken from O'Connell et al (2007). We
plot FUV and NUV AB magnitudes against the unabsorbed power-law X-ray
luminosity in Fig.~\ref{fig:uvbandmag}. We see that there is no
obvious correlation between UV and X-ray fluxes. The FUV and NUV
magnitudes vary from $19-21.5$ and $18-20.5$ respectively, whereas the
X-ray luminosity ranges by almost a factor of 100.

We found only four published values of radio power for the 13
galaxies. These are at 1.4 GHz data and from Miller \& Owen (2001) and
Sijbring (1993). The values of the logarithm of radio power in W
Hz$^{-1}$ are 21.9, 21.2, 21.5 and 21.2 for CGCG 540-101, NGC 1277,
NGC 1278 and NGC 1281, respectively.

\begin{table}
\caption{Total K band flux density and black hole mass of the galaxies}
\label{tab:BHM}
\begin{tabular}{lll} \hline
N& K band flux density & Black hole mass\\
&($10^{-28}$ W m$^{-2}$ Hz$^{-1}$) &($10^8 \Msun$)\\
\hline
1&7.79 &5.58\\
2&4.21 &2.77 \\
3&7.92 &5.69\\
4&14.4 & 11.2 \\
5&3.44&2.22\\
6&2.15&1.30\\
7&3.36&2.16\\
8&6.61&4.66\\
9&1.61&0.94\\
10&6.29&4.31\\
11&0.48&0.24\\
12&4.03&2.95\\
13&3.80&2.48\\
\hline
\end{tabular}
\end{table}

The black hole mass of each galaxy was calculated using the relation
from Marconi \& Hunt (2003), relating $M_{\rm BH}$ and $L_{K,{\rm
    bul}}$ for the `Group 1' galaxies:
\begin{equation}
\log M_{\rm BH}(\Msun)=(8.21\pm0.07)+(1.13\pm0.012)(\log  L_{K,{\rm bul}}-10.9),
\end{equation}
where $L_{K,{\rm bul}}$ is in units of $L_{K\odot}$. We show in
Table~\ref{tab:BHM} the total 2MASS K band luminosities (taken from
NED) and the derived black hole mass.  The masses vary from
$2.4\times10^7$ to $1.1\times10^9\Msun$.

\begin{figure}
  \includegraphics[width=\columnwidth]{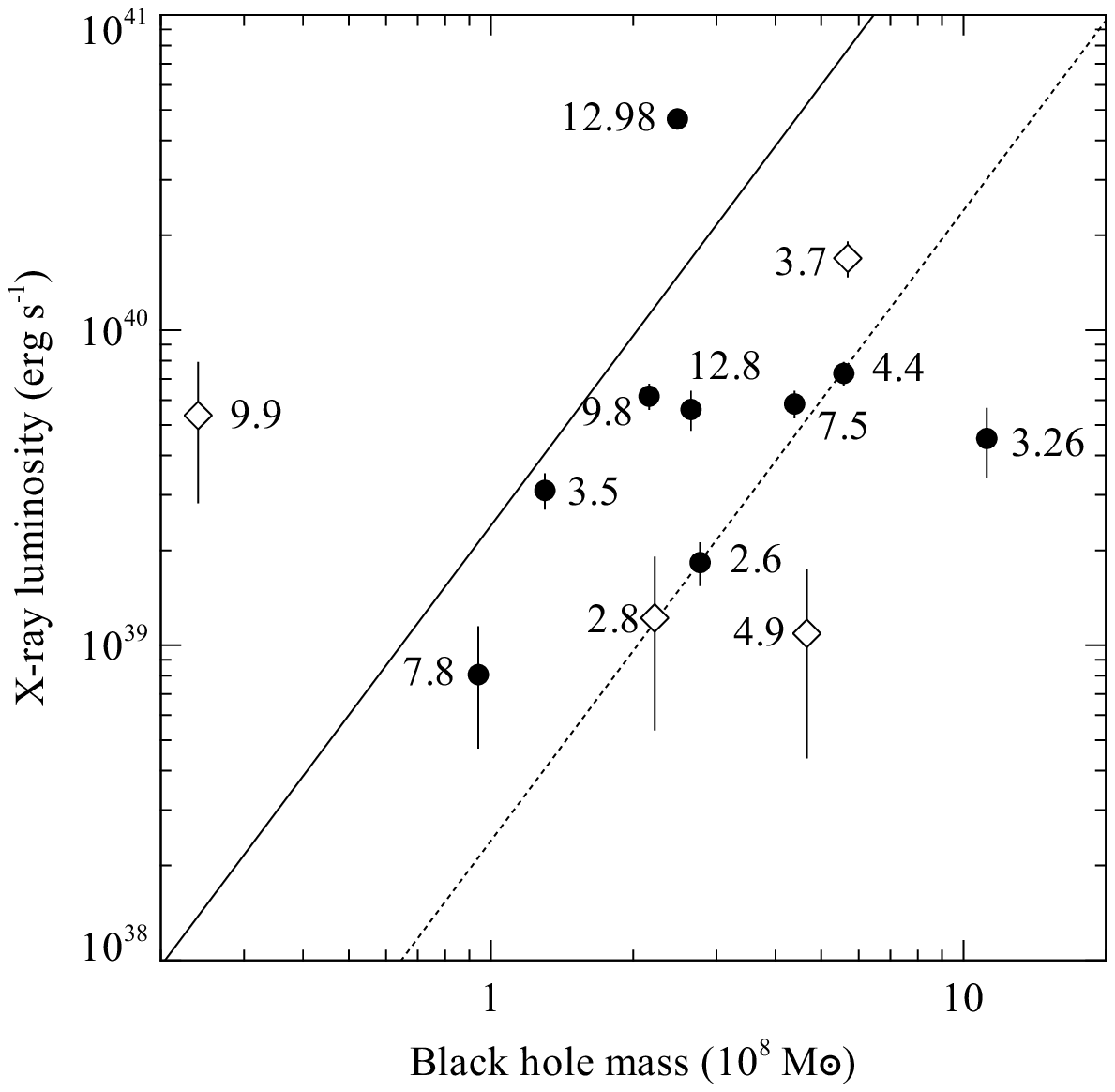}
  \caption{Unabsorbed power-law X-ray luminosity plotted against black
    hole mass. The numbers are the distance in arcmin between the
    central galaxy and the detected galaxies. Objects with a minihalo
    are plotted with an open symbol. The solid line represents the
    expected bolometric luminosity from Bondi accretion and the dashed
    line the 0.5--7~keV luminosity (see Section \ref{sect:discuss}).}
  \label{fig:luminosity}
\end{figure}

The unabsorbed power-law X-ray luminosity was plotted against the
black hole mass of the galaxies in Fig.~\ref{fig:luminosity}. The open
diamonds show those sources with a minihalo. The distances between the
galaxies and the central galaxy, NGC 1275, are also given. We see that
all the sources are within 250 kpc radius from NGC\,1275.  No
correlation is found between the X-ray luminosity and the black hole
mass.

\section{Discussion}
\label{sect:discuss}

All of the bright early-type galaxies in the region studied are
detected in X-rays. The X-ray luminosities range from just below
$10^{39}\ergps$ to $\sim 5\times 10^{40}\ergps$ in an unresolved
component. The sources are spatially coincident, or consistent within
uncertainties, with the centres of the galaxies. The LMXB expected in
these galaxies are barely detectable owing to the high background due
to the dense intracluster gas.

All 13 sources have a power-law spectral component and 4 have an
additional thermal component, suggestive of a mini-halo (as also found
by Sun et al 2007). The photon index of the power-law is steeper than
typically expected from LMXB ($\Gamma\sim 1.6$; Irwin, Athey \&
Bregman 2003), indicating that the sources are unlikely to be central,
unresolved, concentrations of LMXB.

The bolometric accretion luminosity expected from Bondi accretion of
the intracluster medium in the cluster core, assuming that the galaxy
motion is subsonic, is $6\times 10^{40} \left(M_{\rm BH}\over {5\times
    10^8\Msun}\right)^2\ergps$ (e.g. Allen et al 2006). We assume that
the gas has a density of $0.02 \pcmcu$ and velocity/sound speed of
$10^3 \kmps$.  With a typical bolometric correction of 10 in the
0.5--7~keV band for sources at low Eddington rate (Vasudevan \& Fabian
2007), most of our sources have X-ray luminosities close to that
expected from Bondi accretion (Fig.~\ref{fig:luminosity}), especially
when variations in velocity etc. are considered. Where there are
mini-haloes, and it is plausible that they all have some form of
mini-halo due to central stellar mass loss, then the expected
accretion luminosity should be up to $10^3$ times higher owing to the
gas being much denser and cooler.  If these galaxies do contain mini
haloes then they are further extreme examples of the general problem
found for the supermassive black holes in early-type galaxies (Fabian
\& Canizares 1988; Pellegrini 2005). On simple grounds they should all
be much more luminous than observed if the black hole accretes in a
radiatively-efficient manner.  None are as faint as our own Galactic
Centre, Sgr~A*, but there is the fuel at hand for them to be much
brighter.

The solution may lie in ADAFs (Narayan \& Yi 1995) or other
radiatively inefficient flows or with outflows so that little matter
reaches the centre (Blandford \& Begelman 1999). The luminosity of the
sources may also act back on the accreting gas (Ostriker et al. 1976;
Di Matteo et al 2003). Alternatively the matter may accrete to the
centre and power relativistic jets (e.g. Allen et al 2006), which can
be very radiatively inefficient. We do not however see any disturbance
in the surrounding hot gas, nor radio emission, which would be
expected to accompany such powerful jets.

Our results do not distinguish which solution is the more correct, but
underscore the widespread nature of the problem.

\section{Conclusion}
We have made a detailed analysis of X-ray point sources detected with
a deep \emph{Chandra} ACIS-S observation of the core of the Perseus
cluster. The main observational results and conclusions of our study
are:

\begin{enumerate}
\item We have found a total of 13 X-ray sources coincident with the
  nuclei of early-type galaxies projected near the centre of Perseus
  cluster (excluding NGC\, 1275).

\item All 13 sources have a power-law spectra component and 4 have an
  additional thermal component.

\item No obvious correlations are found between X-ray luminosities and
  the K band luminosities, UV AB magnitudes and black hole masses of
  the galaxies.

\item Our results are consistent with the nuclei of all early-type
  galaxies in rich clusters being active, albeit at a low level.
  
\item There is no apparent difference between the X-ray luminosity of
  nuclei in a minihalo compared with those with no detected
  minihalo. Bondi accretion, for those nuclei with a minihalo, should
  make them much more luminous.

\item Some form of radiatively-inefficient accretion is likely
  operating in these sources.
\end{enumerate}

\section*{Acknowledgements}
The authors thank R. O'Connell for the Table of UV fluxes. This
research has made use of the NASA/IPAC Extragalactic Database (NED)
and Sloan Digital Sky survey (SDSS). ACF thanks The Royal Society for
support.

\clearpage

\end{document}